\documentclass{emulateapj}

\shorttitle{First experimental results from the SHARK-VIS forerunner}
\shortauthors{F. Pedichini et al.}

\begin{document}


\title{High Contrast Imaging in the Visible: First Experimental Results at the Large Binocular Telescope}


\author{F. Pedichini\altaffilmark{1,7}, 
M. Stangalini\altaffilmark{1,7},
F. Ambrosino\altaffilmark{1},
A. Puglisi\altaffilmark{2,7},
E. Pinna\altaffilmark{2,7},
V. Bailey\altaffilmark{6},
L. Carbonaro\altaffilmark{2},
M. Centrone\altaffilmark{1},
J. Christou\altaffilmark{3},
S. Esposito\altaffilmark{2,7},
J. Farinato\altaffilmark{5,7},
F. Fiore\altaffilmark{1,7},
E. Giallongo\altaffilmark{1,7},
J. M. Hill\altaffilmark{3},
P. M. Hinz\altaffilmark{4},
and L. Sabatini\altaffilmark{1}}


\altaffiltext{1}{INAF-Osservatorio Astronomico di Roma, 00078 Monte Porzio Catone (RM), Italy}
\altaffiltext{2}{INAF-Arcetri, Florence, Italy}
\altaffiltext{3}{LBTO, University of Arizona, Tucson AZ 85721, USA}
\altaffiltext{4}{CAAO, Steward Observatory, University of Arizona, Tucson AZ 85721, USA}
\altaffiltext{5}{INAF-OAPD, Vicolo dell'Osservatorio 5,35141 Padova, Italy}
\altaffiltext{6}{KIPAC-Stanford University, Stanford CA, 94305 USA}
\altaffiltext{7}{ADONI, INAF ADaptive Optics National laboratory of Italy}

\begin{abstract}
In February 2014, the SHARK-VIS (System for High contrast And coronography from R to K at VISual bands) Forerunner, a high contrast experimental imager operating at visible wavelengths, was installed at LBT (Large Binocular Telescope). Here we report on the first results obtained by recent on-sky tests. These results show the extremely good performance of the LBT ExAO (Extreme Adaptive Optics) system at visible wavelengths, both in terms of spatial resolution and contrast achieved. Similarly to what was done by \cite{amara2012pynpoint}, we used the SHARK-VIS Forerunner data to quantitatively assess the contrast enhancement. This is done by injecting several different synthetic faint objects in the acquired data and applying the ADI (angular differential imaging) technique. A contrast of the order of $5 \times 10^{-5}$ is obtained at $630$ nm for angular separations from the star larger than $100$ mas. These results are discussed in light of the future development of SHARK-VIS and compared to those obtained by other high contrast imagers operating at similar wavelengths.   
\end{abstract}


\section{Introduction}
LBT and, more specifically, its FLAO (First Light Adaptive Optics) system \citep{Esposito2010, esposito2010first, quiros2010first, esposito2012natural, 2014SPIE.9148E..03B}, recently opened a new frontier for the astronomical AO on the 8-10 m class telescopes, by routinely delivering strehl ratios (SRs) higher than $0.8$ in H band. This led to important scientific breakthroughs \citep{esposito2013lbt, skemer2012first}. The combination of the pyramid wavefront sensor, together with the high dynamic and spatial resolution of the ASM (adaptive secondary mirror), provide performance never obtained on this class of telescope by previous natural nor laser guide star systems. In particular, we refer to the extremely low residual wavefront error (below $100$ nm rms), reached by the FLAO systems working with bright guide stars ($R<9.5$) in good seeing conditions (below $0.8$ arcsec). These constitute very promising conditions for extending the operational range of the LBT AO to visible wavelengths. \\
In this regard it is worth mentioning that there are several scientific advantages in working in the visible band \citep{2014SPIE.9148E..1MC}. First, visible detectors are less noisy and more linear. In addition, they are characterized by a larger dynamic range and are easier to operate than the current generation of NIR (Near-Infra-Red) ones. Second, visible skies are much darker than ones in the NIR bands. In addition from the scientific point of view, we remark that most of the strongest emission lines are in the visible bands (i.e. H$\alpha$ line). Moreover visible AO systems have a higher spatial resolution, up to a factor of $\sim{3}$, than AO systems working in K-band. Furthermore, models of exoplanet atmospheres \citep{marley1999reflected, fortney2008unified, marley2011probing}  show that at wavelengths shorter than $650$ nm the planet albedo increases, so that the probability of their detection is maximized. More recently, a large effort has been put to the design of two high contrast imagers at LBT, SHARK-VIS and SHARK-NIR, exploiting FLAO at visible and infrared wavelengths, respectively \citep{Farinato2014_referred, jacopo2014, stangalini2014solar}.
\begin{figure}[ht]
   \centering
   \includegraphics[width=7cm, clip]{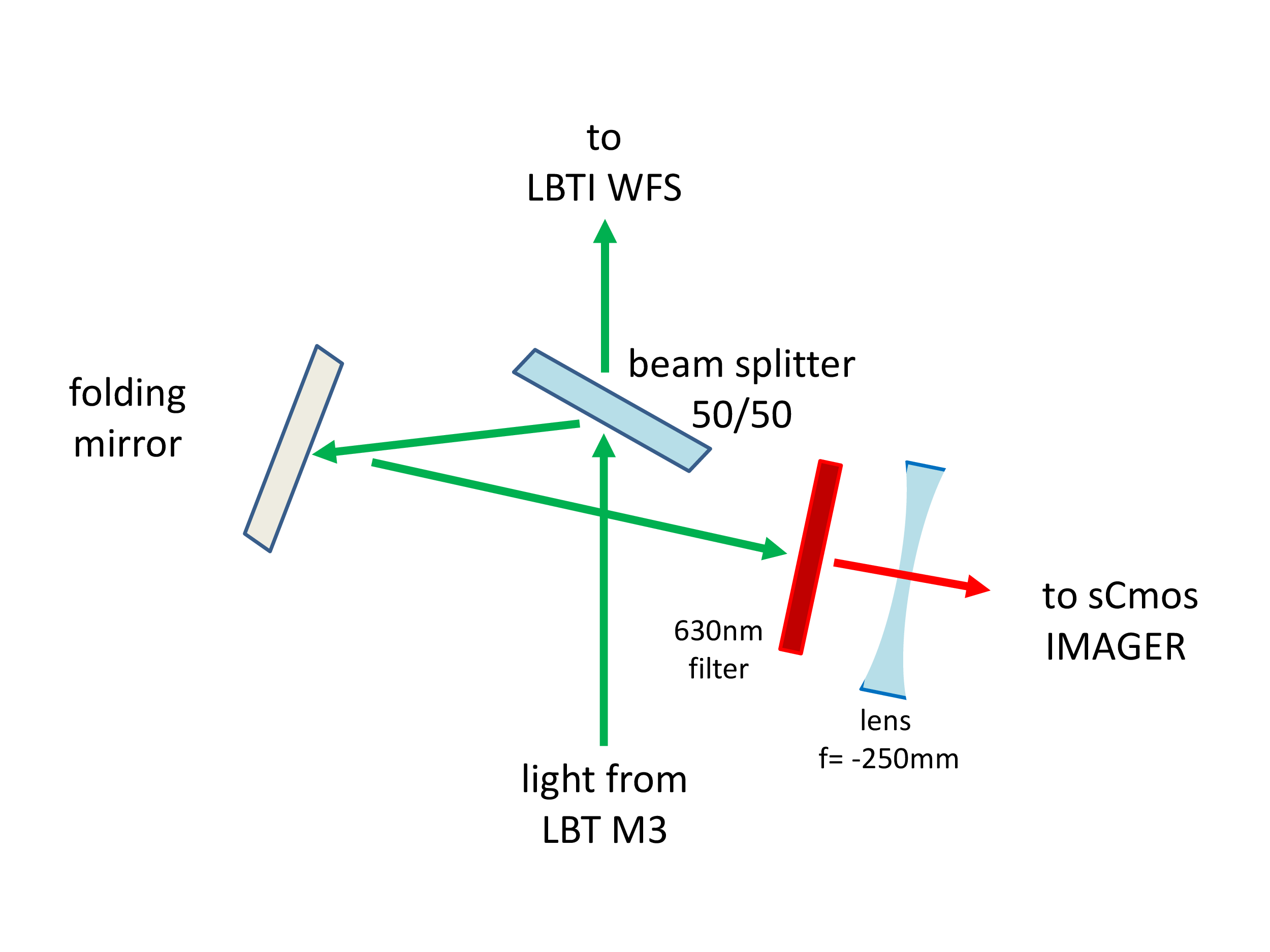}
     \caption{The SHARK-VIS Forerunner optical layout} 
    \label{schematics}
   \end{figure}
These imagers have been conceived to minimize NCPA (non-common-path aberration). Indeed, they are positioned very close to the pyramid wavefront sensor and minimize the number of optical surfaces employed.
High contrast imaging is nowadays becoming attractive, with a number of recent instruments optimized for both visible wavelengths (VisAO,  \cite{2014ApJ...786...32M,2014SPIE.9148E..1MC,2014IAUS..299...46M}) and NIR bands (i.e. GPI, \cite{2014PNAS..11112661M, 2014SPIE.9149E..2BR,2008SPIE.7015E..18M} and SPHERE, \cite{2015ApJ...800L..24H,2006Msngr.125...29B,2015ESS.....320305B}). 
In this paper, we describe our recent results obtained on-sky with the SHARK-VIS experimental imager, hereafter Forerunner. In order to estimate the Forerunner contrast, we injected faint synthetic targets into the acquired data at different angular separations from the bright source, and we applied ADI post-processing technique. This approach was already proposed and employed by \citet{amara2012pynpoint} for testing different post-processing techniques and assessing their contrast enhancement. Such an approach also represents an accurate way to evaluate the performance of the instrument with on-sky data, thus without making use of complex, yet not completely realistic, numerical simulations.
   \begin{figure*}[h!]
   \begin{center} 
   \includegraphics[width=7.5cm, clip]{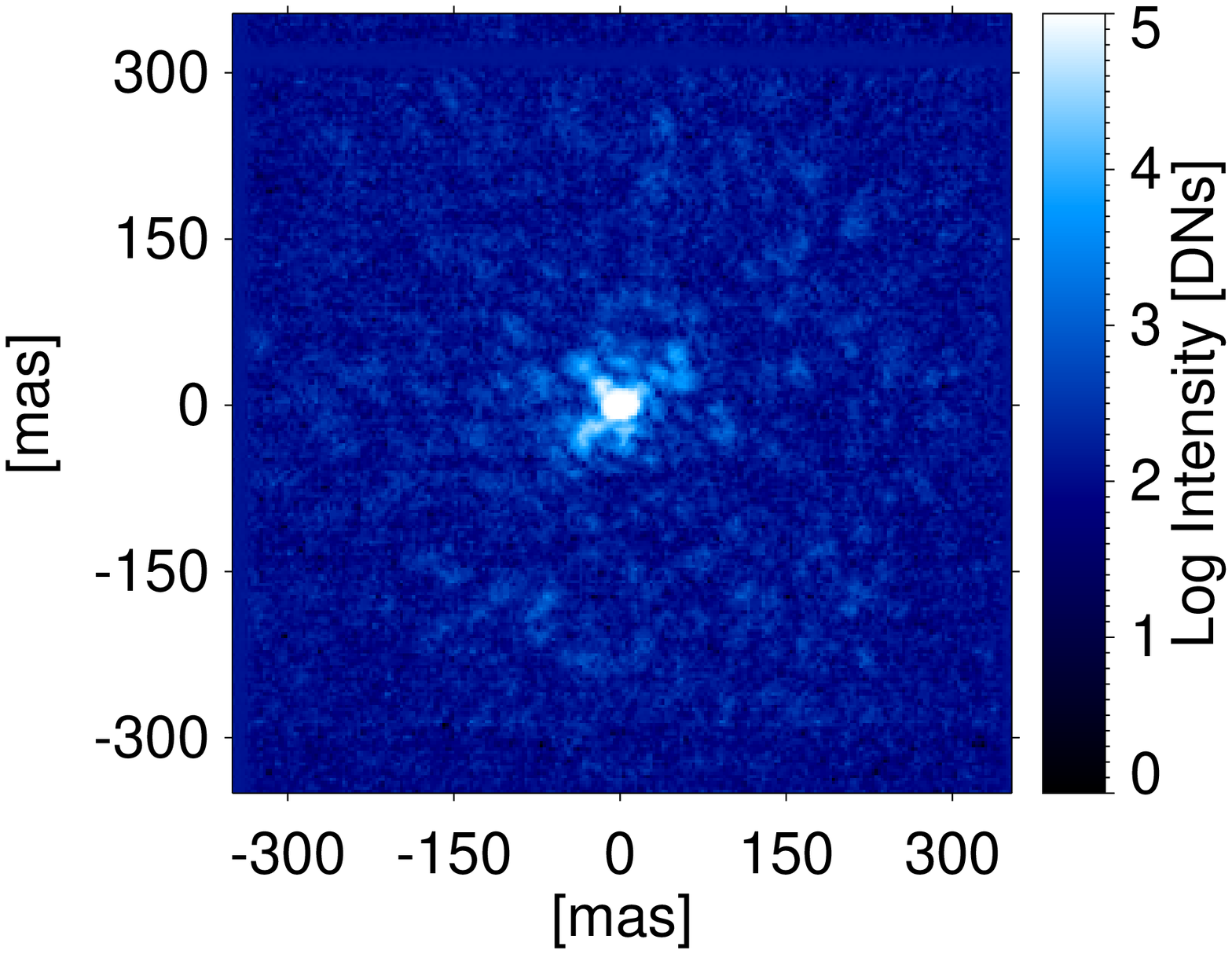} 
   \includegraphics[width=7.5cm, clip]{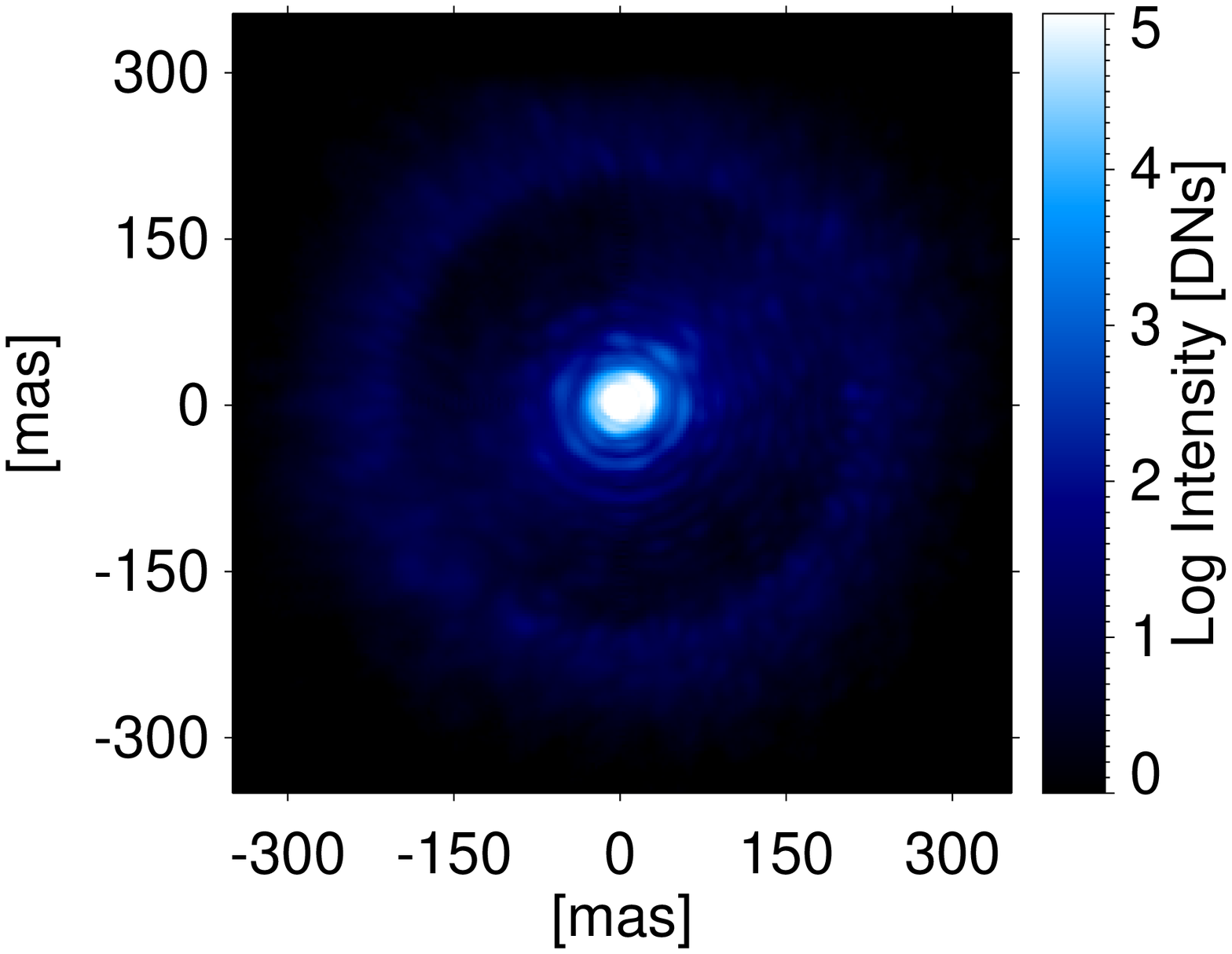} 
   \includegraphics[width=8cm, clip]{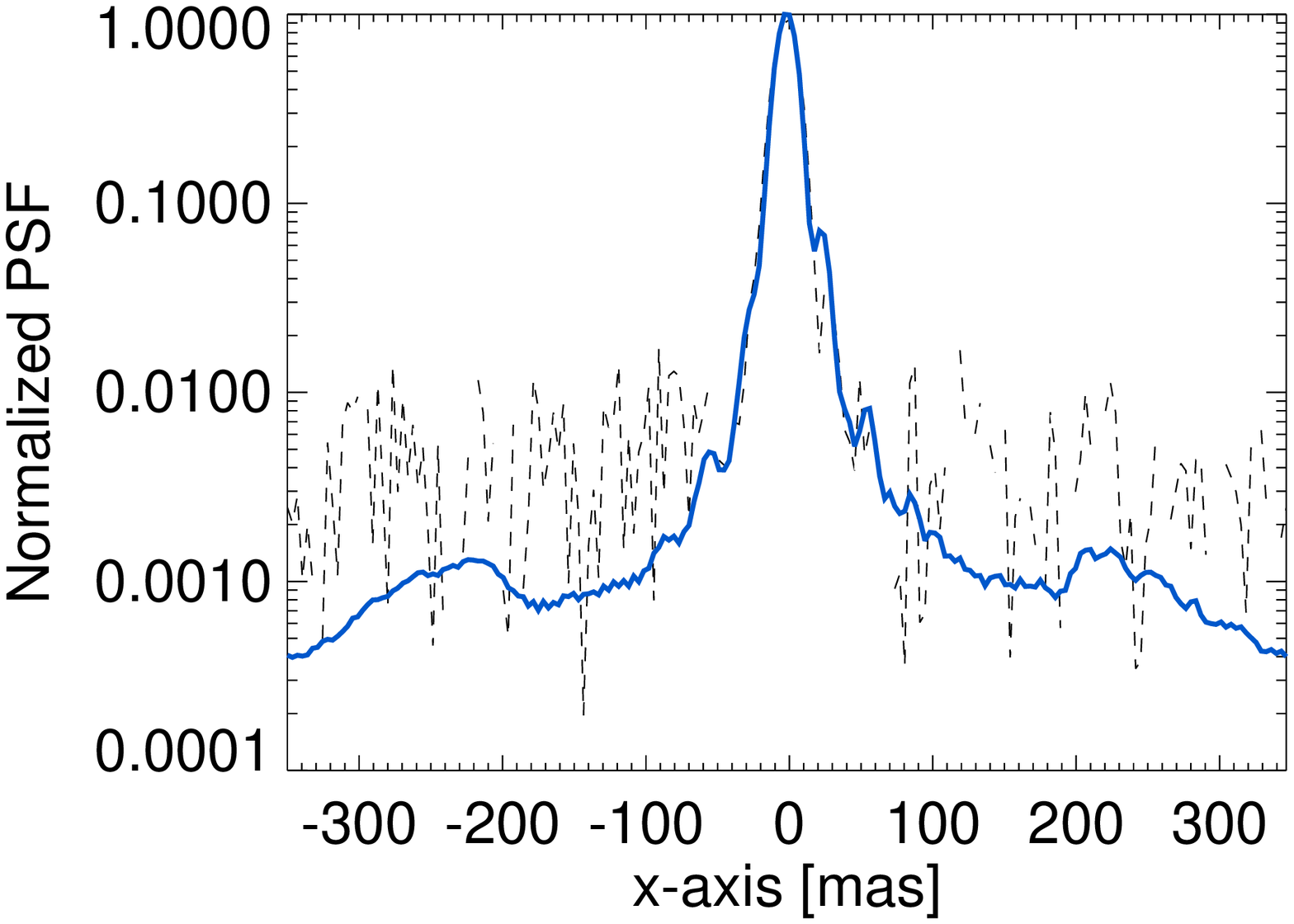} 
   \end{center}
   \caption[sequence] 
   { \label{fig:imgs} 
Top left: short exposure ($1$ ms) PSF of GLIESE 777. Top right: long exposure ($5$ s) PSF of GLIESE 777. Bottom: radial profile of both short (dashed line) and long (blue continuous line) exposure PSFs.}
\end{figure*} 

 \section{The Forerunner}
The Forerunner was installed at the right bent Gregorian focus of LBT in February, 2014, and finally tested on sky in June, 2015. The simplicity of this instrument (see Fig.~\ref{schematics}), together with the ASM (Adaptive Secondary Mirror) of LBT, reduces the number of refractive optical surfaces between the sky and the detector to only two optical elements. The first one corresponds to an interference filter centered at $630$ nm with a bandwidth of $40$ nm, and the other is a divergent lens of $250$ mm of focal length used to achieve a slight Nyquist oversampling of the PSF (Point Spread Function), with a spatial scale of $3.73$ mas/pixel. The filter bandwidth has been limited to only $40$ nm because of the lack of an ADC (Atmospheric Dispersion Corrector) in the optical layout of this basic pathfinder experiment. This particular configuration allows operation only up to $10$ degrees of Zenith distance. However, this limitation will be overcome in the final design of SHARK-VIS, where an ADC is foreseen.\\
NCPAs are reduced by positioning the detector close, and mechanically connected, to the LBTI (LBT Interferometer) main frame that is holding the pyramid wavefront sensor of the AO system. Small residual static aberrations on the science camera ($\sim 15$ nm) were minimized by offsetting the zero point values of low order modes of the wavefront reconstructor. The camera used is an Andor Zyla\footnote{http://www.andor.com/scientific-cameras/neo-and-zyla-scmos-cameras/zyla-55-scmos} hosting a sCMOS sensor cooled to $0^{\circ}$ Celsius. This camera can perform high speed imaging acquisition with low noise ($\le 1e^-/pixel$). In fact, the deployed system allows recording $2k \times 2k$ pixel images at $50$ Hz, and $200 \times 200$ pixel subfields (the format used in this experiment) at $1$ kHz. \par
In Fig. \ref{fig:imgs} we show examples of both a short exposure ($1$ ms) image with its relative radial profile, and a long exposure ($5$ s) image (obtained by co-adding $5000$ images), with its own radial profile after subtracting the dark frame from the images and co-registering the data series by means of a FFT (Fast Fourier Transform) phase correlation technique. Several diffraction rings and the control radius are also evident in the ($1$ ms) image. The control radius, which is evident in the long exposure of Fig. \ref{fig:imgs} (right upper panel) as an annulus of increased brightness at $\approx$ 240 mas from the central peak of the PSF, marks the region within which the action of the AO takes place. Its radius, which depends on the number of actuators (on the telescope pupil), is in perfect agreement with the theoretical value of $ 206265 \times N_{act} \times \lambda /D = 0.238 $ arcsec, where $N_{act}=15$ is the number of actuators on the LBT pupil radius and $D=8.2$ m is the effective pupil diameter set by the undersized secondary mirror size.

\section{Data set}
\begin{figure*}[t]
   \begin{center}
   \includegraphics[height=5.5cm]{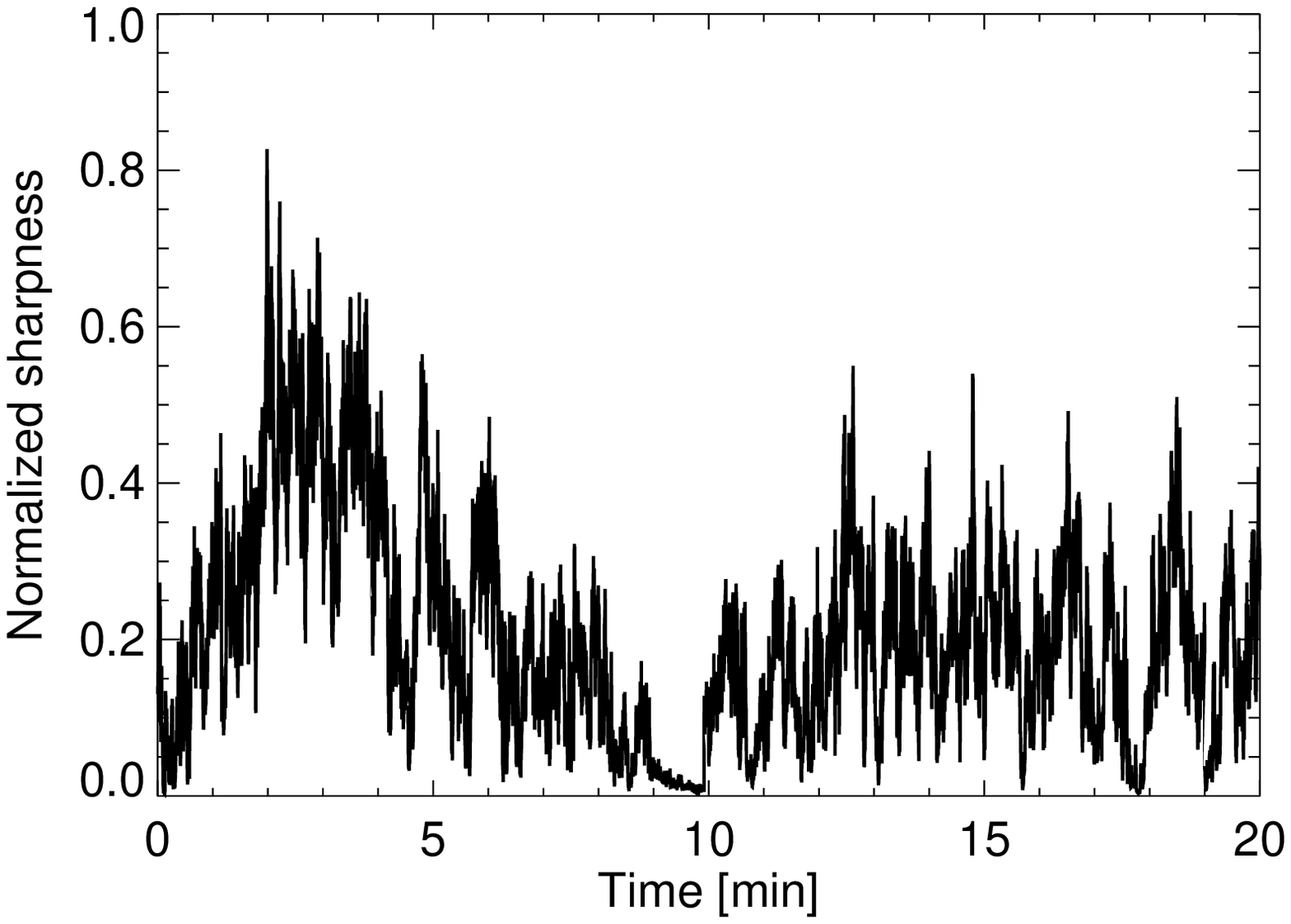}
   \includegraphics[height=5.5cm]{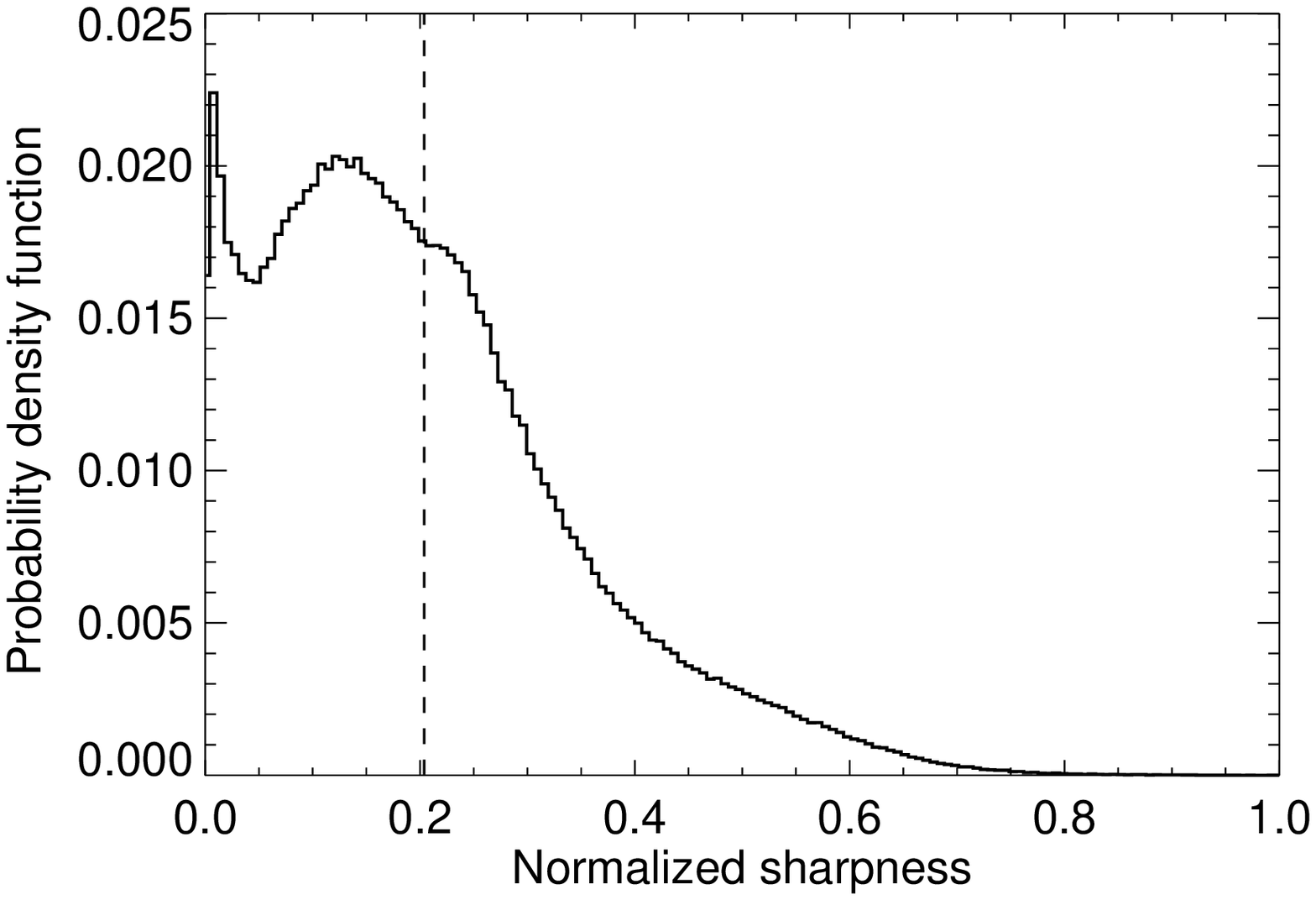}
   \includegraphics[height=5.5cm]{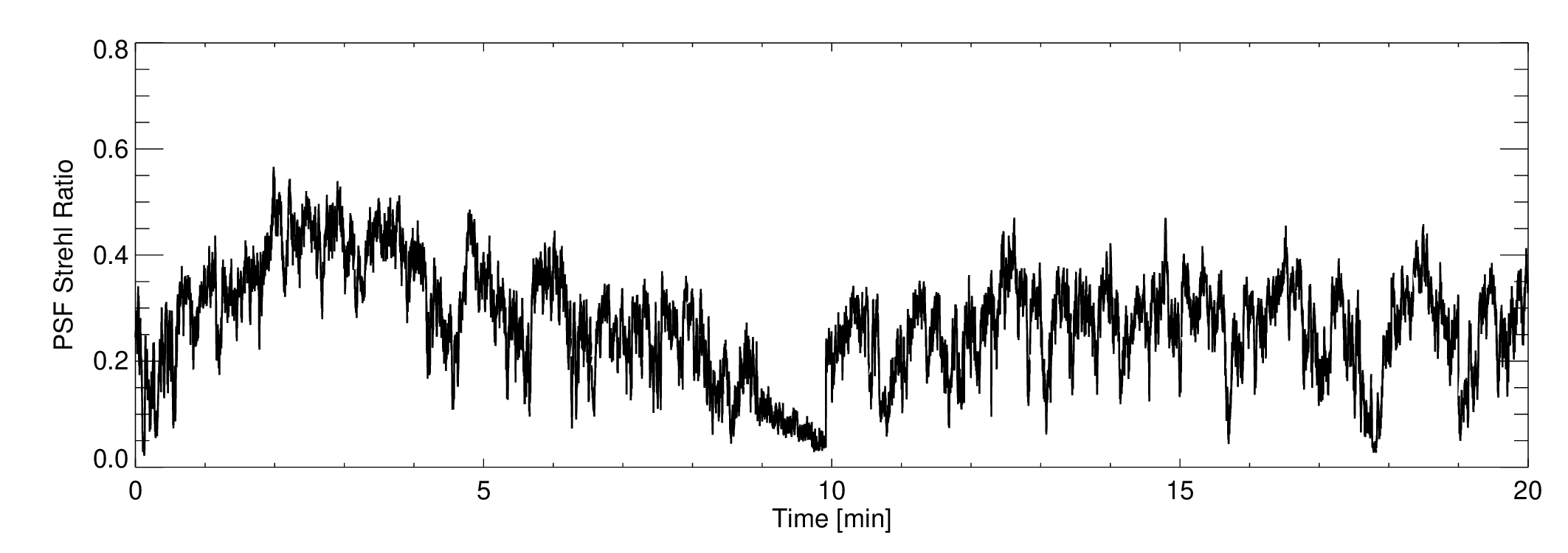}
   \end{center}
   \caption[example] 
   { \label{fig:intensitypeak} 
Upper left: Sharpness, normalized to its maximum, as a function of time. In order to reduce the number of data points for graphical reasons, an averaging window of 100 samples was applied to the data. Upper right: Probability density function of the sharpness. The distribution displays a large dispersion, which reflects the variations of the seeing during the observation. In addition, the PDF also shows different populations which again manifest different conditions. Bottom: PSF SR as a function of time, for the complete dataset.}
\end{figure*}

The data set we present in this paper is composed of a sequence of 1,200,000 images of the target Gliese $777$ recorded at $1$ ms cadence ($200 \times 200$ pixel subfields) on June $4^{th}$ 2015 starting from 08:21:58 UT. The LBT AO system was correcting $500$ modes in closed loop, while seeing conditions were rapidly varying in the range $0.8-1.5$ arcsec. 
This can be seen in Fig.\ref{fig:intensitypeak} (upper left panel) where we plot the evolution of the sharpness of the images, normalized to its maximum,  $vs.$ time. Here we have used the same definition of sharpness as in \citet{muller1974real}:
$$S = \int dx dy I^{2} (x,y),$$
where $x$ and $y$ denote coordinates in the image plane and $I$ is the intensity. In the upper right panel of the same figure we plot the PDF (Probability Density Function) of the sharpness. This distribution shows the presence of at least two peaks, which manifest different seeing conditions during the observation. In the bottom panel of Fig.\ref{fig:intensitypeak} we also show the SR behaviour. We note that the SR undergoes rapid fluctations ranging from a few percent up to $50\%$. Its average value over the whole duration of the observation is $27\%$.
A residual tip-tilt with an rms amplitude of $\sim 17$ mas and main frequency of $\sim 13$ Hz has been found in the raw data, probably due to wind excited modes of the secondary and tertiary mirror spider supports. The Forerunner has no image rotator, hence the image de-rotation is performed by post processing of short sub-exposures.
It is worth stressing that the high frame rate of the data allows the post-facto minimization of the residual tip-tilt. Indeed, this would have not been possible using longer integration times.

\section{Methods and Results}

Here we focus our attention to the assessment of the performances of the SHARK-VIS Forerunner in terms of contrast. 
Using on-sky data as a realistic benchmark for estimating contrast capabilities of the instrument has been already done by \cite{amara2012pynpoint}. Following the latter work, we inject in each frame of our data sequence faint objects at different radial distances from the central source with flux ratios of $10^{-4}$ and $5 \times 10^{-5}$. This is done by re-scaling the instantaneous PSF and adding it in the image, at different separations from the central object.

The data reduction strategy based on the ADI concept can be summarized as follows:
\begin{enumerate}
\item subtracting the dark frame; 
\item co-registering images through a FFT phase correlation technique. This registration method, being applied in the Fourier space, allows the registration of the whole series of images with sub-pixel accuracy, thus minimizing the residual tip-tilt error; 
\item injecting synthetic faint objects, for each one of the sub-frames (at different orientations reflecting the parallactic angle and thus the field rotation), with a shifted and amplitude re-scaled PSF. These objects have been placed at different distances from the optical axis (i.e. $45$ mas, $95$ mas, $190$ mas, and $290$ mas) and their fluxes correspond to different contrast ratios with respect to the peak value of the bright source at the center of the field (i.e. $10^{-4}$, and $5 \times 10^{-5}$); 
\item applying ADI technique \citep{marois2010images, vigan2010photometric} on the data. For each one of the frames the bright star PSF is removed subtracting its estimate computed by the median operator over $5000$ random frames selected throughout the overall sequence. Then the residuals of each frame are de-rotated and combined with a median operator to retrieve the final image. This is finally flatted removing low spatial frequencies estimated using a boxcar median operator with 11 pixel width.
\end{enumerate}
It is worth noting here that large variations of the seeing during the observation may lead to inaccuracies in the subtraction of the PSF from each time frame; as this is obtained as a median over time of evolving PSFs.

\begin{figure*}
   \begin{center} 
   \begin{tabular}{c}
   \includegraphics[trim={0.8cm 0 4cm 2},clip,width=0.43\textwidth]{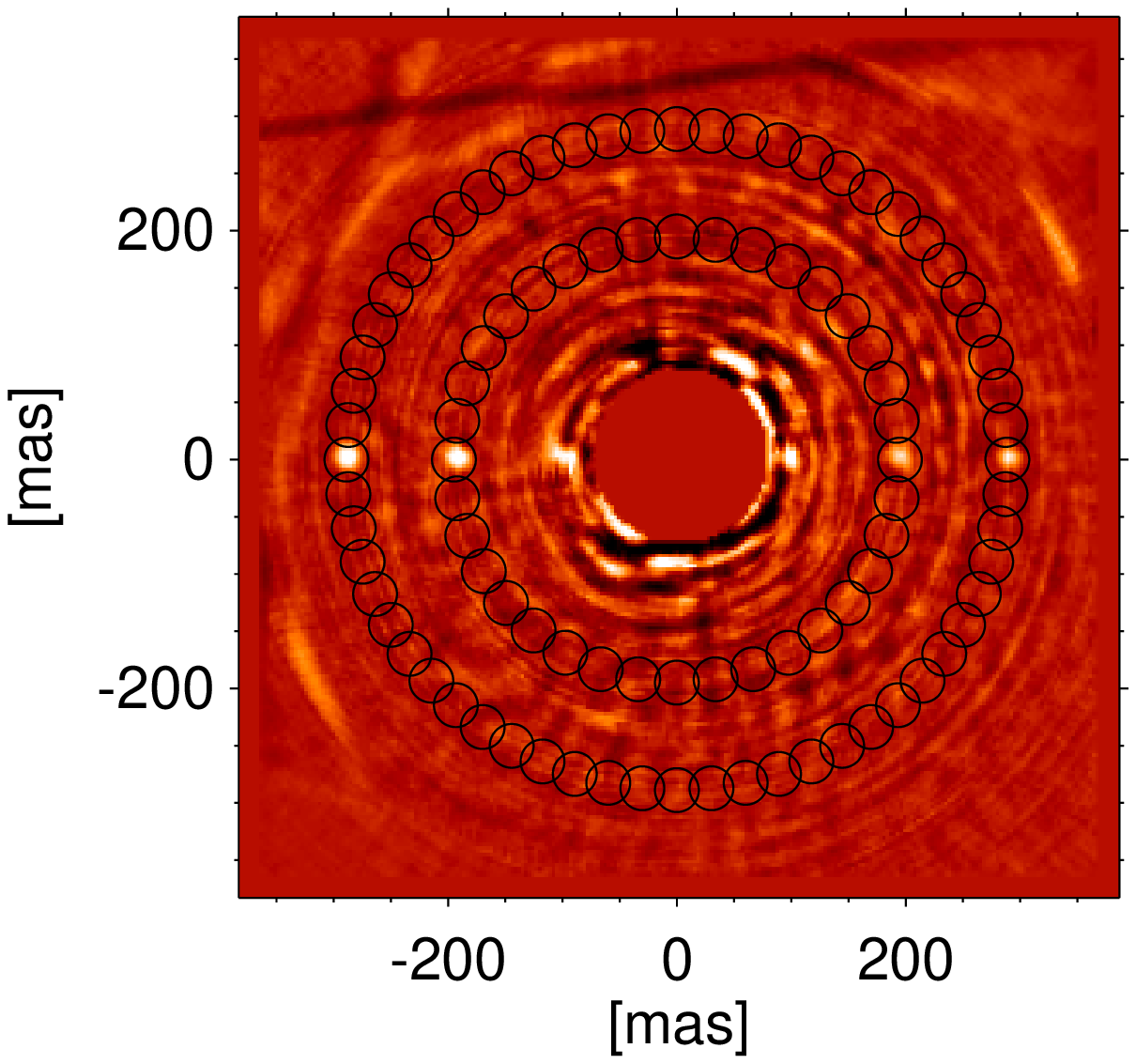} 
   \includegraphics[trim={1.2cm 0 0 0}, clip,width=0.55\textwidth]{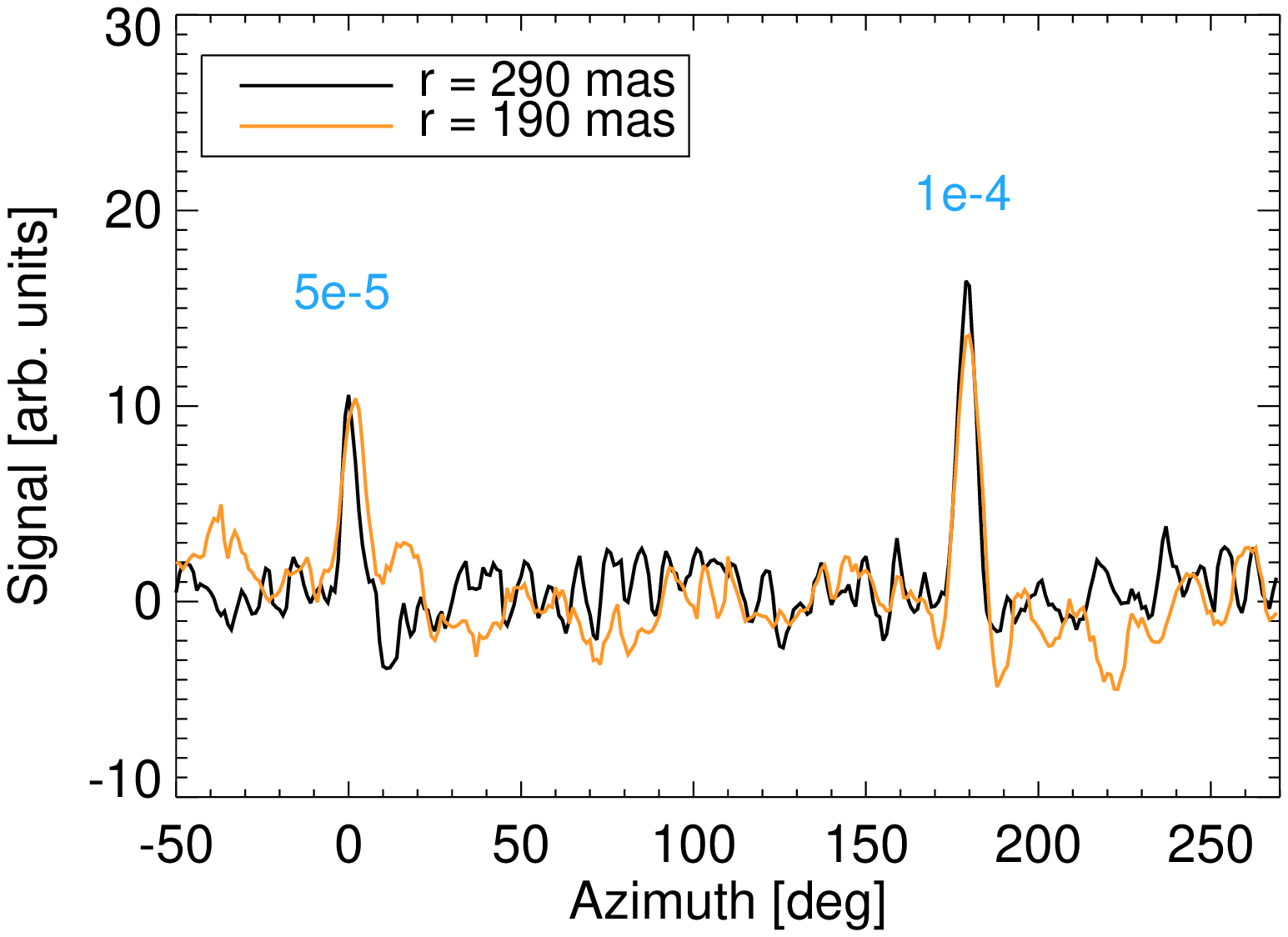}\\
   \end{tabular}
   \end{center}
   \caption[sequence] 
   { \label{fig:SN} 
Left: ADI detection map after removing a background map estimated through a $11$ pixel width median filter. The black circles represent some of the apertures used to measure the photometric signal of fake planets. Right: photometry measurements over the circular apertures shown in the left panel; the two peaks represent the detection of the fake planets at contrast of 5e-5 and 1e-4.}
\end{figure*}

In the left panel of Fig. \ref{fig:SN} we show the final ADI detection map obtained by following the procedure described above and representing the final result of 1200 seconds of exposure on GLIESE 777 without discarding any image during the post processing. All planets located beyond $100$ mas from the host star are well detectable, although they have different S/N ratios due to different contrast ratios ($10^{-4}$, and $5 \times 10^{-5}$) as well. Planets located closer to the bright source are comparable to the noise level so they can't be well detected. 
In order to compare the achieved contrast inside and outside the AO control radius, in table \ref{SN_table} we report the fluxes and S/N of the photometry of fake planets injected at 190 and 290 mas, respectively. The flux values are measured using circular apertures of 4 pixel radius and the local sky level, to be subtracted, is measured inside an annulus of respectively 5 and 7 pixel radii. We estimate the noise of these measures as the RMS value of a set of similar photometric estimates taken away from the "planet" location, but distributed along the same radial distance, as shown in the left panel of figure \ref{fig:SN}. Other more sophisticated photometry algorithms may provide better results but we prefer the use of this very conservative approach for the assessment of our detection limit.

\begin{table}[ht]
\centering
\caption{Fake planets fluxes and their S/N vs radial distances}
\label{SN_table}
\begin{tabular}{lllll}
\hline
mas & \begin{tabular}[c]{@{}l@{}}1e-4\\ flux\end{tabular} & \begin{tabular}[c]{@{}l@{}}1e-4\\ S/N\end{tabular} & \begin{tabular}[c]{@{}l@{}}5e-5\\ flux\end{tabular} & \begin{tabular}[c]{@{}l@{}}5e-5\\ S/N\end{tabular} \\ \hline
190 & 13.4                                                & 8.1                                                & 10.0                                                & 5.9                                                \\
290 & 17.8                                                & 15.1                                               & 9.5                                                 & 8.1                                              
\end{tabular}
\end{table}
In the right panel of Fig. \ref{fig:SN}, we plot the photometric signal measured in apertures at two constant radial distances (see left panel of he same figure) as a function of the azimuth. The radial distances chosen are $190$ and $290$ mas, corresponding to the distance of the injected planets, within and outside the control radius of the AO. Please note that, for graphical reasons, the number of apertures plotted in the left panel of Fig. \ref{fig:SN} was largely reduced with respect to the exact number used to estimate the signals shown in the right panel of the same figure.

\section{Concluding remarks}
We have used the first on-sky data acquired by the Forerunner as test data for an accurate assessment of the overall performances of the instrument in terms of contrast. To this aim, we have injected synthetic faint sources (a.k.a. fake planets) into the data at different distances from the optical axis, and have applied the ADI  post-facto technique to estimate the contrast enhancement achieved. As noted by \cite{amara2012pynpoint} using on-sky data offers several advantages over the use of end-to-end simulations since the residual wavefront aberrations and any possible quasi-static speckle induced by the instrument, are the real ones. This allows an accurate assessment of the performance of the instrument without making use of complex, yet not fully realistic, numerical simulations.\
Our analysis shows that the Forerunner can reach contrast down to $5 \times 10^{-5}$, with a good S/N at radial distances greater than $100$ mas. This is an excellent result considering that it is obtained at visual wavelengths, where the effects of residual seeing aberrations are stronger than at longer ones.
Our results are comparable with those obtained by the VisAO instrument which operates at similar wavelengths ($0.63-1.05\,\mu m$), even if they can reach a contrast of $10^{-5}$ at the different angular separation of $500$ mas, but exploiting better seeing conditions ($\sim 0.65$ arcsec) and with a frame selection of 50\% over a long exposure of $5$ h  \citep{2014ApJ...786...32M}. 
In any case, we want to stress that our results lead to very interesting frontiers in direct imaging of exoplanets considering following aspects:
\begin{itemize}
\item the Forerunner has not yet made use of a coronagraphic system;
\item seeing was rapidly varying during the acquisition ($0.8-1.5$ arcsec);
\item limited total exposure time (only $20$ minutes);
\item no frame selection during ADI post processing;

\end{itemize}
Considering that our data sequence is much shorter ($20$ min) than the one used to assess the performance of VisAO itself (2.5 hrs) and our seeing conditions were much less favorable than those occurred during the on-sky test of VisAO the Forerunner can be regarded as a very promising experimental instrument.
This is especially the case if one considers the foreseen upgrades SOUL (Single conjugated adaptive Optics Upgrade for LBT) of the LBT AO system and the future SHARK-VIS instrument for high contrast imaging in the Visual bands at LBT. For more information about SOUL we refer the reader to \cite{soul}.

\section{Acknowledgments} 
The LBT is an international collaboration among institutions in the United States, Italy and Germany. LBT Corporation partners are: The University of Arizona on behalf of the Arizona Board of Regents; Istituto Nazionale di Astrofisica, Italy; LBT Beteiligungsgesellschaft, Germany, representing the Max-Planck Society, the Leibniz Institute for Astrophysics Potsdam, and Heidelberg University; The Ohio State University, and The Research Corporation, on behalf of The University of Notre Dame, University of Minnesota and University of Virginia.\\
This work was partially funded by ADONI, the ADaptive Optics National laboratory of Italy.\\


\end{document}